\documentclass[aps,prl, twocolumn,groupedaddress,amsmath,ams-fonts,showpacs,dvipdfmx]{revtex4-1}

\usepackage{graphicx}
\usepackage{dcolumn}
\usepackage{bm}
\usepackage{here}
\usepackage{color}
\usepackage{relsize}

\usepackage{braket}

\usepackage{BOONDOX-frak}

\def \beal{\begin{align}}

\def \enal{ \end{align} }

\def \part{\partial}

\def \vareps{\varepsilon}

\def \lam{\lambda}

\def \sig{\sigma}

\def \Lam{\Lambda}
\def \nab{\nabla}

\begin{document}

\title{Torsional chiral magnetic effect in Weyl semimetal with topological defect}

\author{Hiroaki Sumiyoshi$^1$}
\author{Satoshi Fujimoto$^{2}$}%
\affiliation{$^1$Department of Physics, Kyoto University, Kyoto 606-8502, Japan}
\affiliation{$^2$Department of Materials Engineering Science, Osaka University, Toyonaka 560-8531, Japan}

\date{\today}

\begin{abstract}
We propose a torsional response raised by lattice dislocation in Weyl semimetals akin to chiral magnetic effect;
i.e. a fictitious magnetic field arising from screw or edge dislocation induces charge current.
We demonstrate that, in sharp contrast to the usual chiral magnetic effect which vanishes in real solid state materials, the torsional chiral magnetic effect exists even for realistic lattice models, which implies
the experimental detection of the effect via SQUID or nonlocal resistivity measurements in Weyl semimetal materials.
\end{abstract}

\maketitle
Recently, many candidate materials for Dirac semimetals and Weyl semimetals (WSMs) \cite{vishwanath,balents,balents2,murakami}, have been discovered \cite{Kim_BiSb2013,Zhijun2012, Zhijun2013, Liu_NaBi2014,Neupane2014, Borisenko2014, Liu_CdAs2014,Li_ZrTe2014,TaAsWeng2015, Huang_TaAs2015, Xu_TaAs2015, Zhang_TaAs2015,LvTaAs2015, Shekhar2015,XiaochunTaAs2015, Yang_NbAs2015,Shekhar_TaP2015,Sushkov_EIO2015,Borisenko}. 
These topological semimetals are intriguing because of
exotic transport phenomena 
associated with the chiral anomaly in quantum field theory 
\cite{Fujikawa},
such as  
the anomalous Hall effect \cite{Zyuzin_AHE2012, Goswami2013},
chiral magnetic effect (CME) \cite{Hosur_review},
negative longitudinal magnetoresistance \cite{NN_1983, Kim_BiSb2013, Li_ZrTe2014,Yang_NbAs2015, Shekhar_TaP2015, Goswami_LMR2015, XiaochunTaAs2015}, and
chiral gauge field \cite{Liu2013}.

Among them, the CME has been discussed in broad areas of quantum many-body physics, including nuclear and nonequilibrium physics as well as condensed matter physics.
It is the generation of charge current parallel to an applied magnetic field even in the absence of electric fields.
In nuclear physics, together with the chiral vortical effect \cite{CVE2010}, it is expected to 
play an important role 
in heavy ion collisions experiments \cite{Kharzeev2014, Fukushima_CME2008}.
The CME also caused a stir in nonequilibrium statistical physics,
since it leads to the existence of the ground state 
which, recently, attracts a renewed interest
in connection with the realization of quantum time crystal
\cite{Wilczek2012},
and then the CME has been studied from this point of view \cite{Vazifeh2013, Yamamoto_2015}.
However, unfortunately, their results are negative for its realization:
the macroscopic ground state current in realistic WSMs is always absent.




In this letter, we propose a chiral response in WSMs, named ``torsional chiral magnetic effect (TCME)",
in which the ground state charge current is caused by the effective magnetic field induced by lattice dislocation
as shown in FIG.\ref{current}.
By using the Cartan formalism of the differential geometry, we can describe
the lattice strain and dislocation in terms of vielbein and torsion \cite{previous-study}.
From the viewpoint of the quantum field theory in curved space-time, the TCME is raised by the mixed action of electromagnetic and torsional fields that is
prohibited in four-dimensional spacetime with the Lorentz symmetry,
but made possible in non-relativistic band electrons in solid state systems.
Furthermore, we demonstrate that the TCME is possible in realistic lattice models by carrying out numerical calculations.
Our results imply the existence of experimentally observable current induced by the TCME in real WSM materials.
We also resolve the relation between our results and the no-go theorem that the CME is absent in equilibrium states \cite{Vazifeh2013, Yamamoto_2015}.
First of all, we clarify the notations. 
The indices $i,j,\cdots=x,y,z$ and $a,b,\cdots=\bar{x},\bar{y},\bar{z}$ represent the coordinates in the laboratory and local orthogonal (or Lorentz) frames, respectively.
In the following, we use the Einstein summation convention.



{\it Linear response theory for torsional response---}
Here, we briefly introduce the Cartan formalism, which can be applied to description of crystal systems with lattice strain as follows.
It is an approach to curved space 
and based on the local orthonormal frame form $ e^a=e^a_i({\bm r})dr^i $,   
where the coefficient fields $e^a_i({\bm r})$ are referred to as the vielbein \cite{Nakahara_topo}.
We introduce the coordinate measured by an observer on the deformed lattice $R^{a}$ and the laboratory coordinate $r^i$, and
identify its exterior derivative as the local orthonormal frame $e^a=dR^a$.
Then, to the first order in the displacement field $\vec{u}$, the vielbein is written as $e^a_i=\delta^a_i-\part{u^a}/\part{r^i}
\begin{figure}
 \begin{center}
\includegraphics[width=80mm]{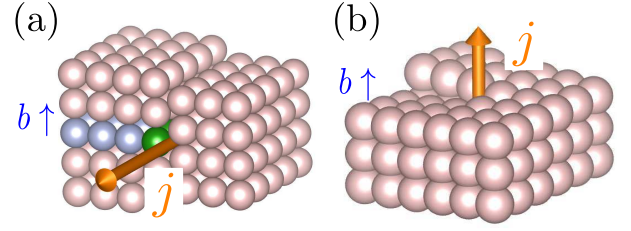}
 \end{center}
\caption{
(Color Online)
Ground state current $j$ induced by (a) edge and (b) screw dislocation
with the Burgers vector $b$.
 }                                                                                                                                                                                                                                         
 \label{current}
\end{figure}

$.
For this observer, the lattice is not deformed,
and then the Hamiltonian of the system is given by
$H(-\mathrm{i}\part_{R^{\bar{x}}},-\mathrm{i}\part_{R^{\bar{y}}},-\mathrm{i}\part_{R^{\bar{z}}}) = H(-\mathrm{i}e^i_{\bar{x}}\part_{r^i},-\mathrm{i}e^i_{\bar{y}}\part_{r^i},-\mathrm{i}e^i_{\bar{z}}\part_{r^i})$,
where $H(p_x,p_y,p_z)$ is the Hamiltonian without lattice deformation and $e^{\mu}_{\alpha}$ is the inverse of $e_{\mu}^{\alpha}$.
In this way, the emergent vielbein appears, and therefore we can describe the elastic response by using the Cartan formalism.
The coupling between the vielbein and electrons is similar to the minimal coupling of the $U(1)$ gauge field, $p_i \to p_i - e A_i$.
Then, we can define the analog of the field strength by $T^{a}_{ij}=\part_{i}e^{a}_{j}-\part_{j}e^{a}_{i}$,
which is referred to as the torsion, or ``torsional magnetic field" (TMF) \cite{Shitade2014,Parrikar2014,torsional_mag},
where the spin connection is dropped for simplicity.
Using the displacement vector, the torsion is rewritten as $T^a_{ij}=(\part_{j}\part_{i}-\part_{i}\part_{j})u^a$.
The point is that, if $u^a({\bm r})$ is a well-defined function, the torsion is always zero,
and the multivaluedness of $u^a({\bm r})$ is necessary for nonzero torsion.
Indeed, the edge dislocation along $z$-axis with Burgers vector $b_g\hat{y}$ causes the TMF, $T^y_{xy}=-b_g\delta^{(2)}(x,y)$,
and the screw one with $b_g\hat{z}$, $T^z_{xy}=-b_g\delta^{(2)}(x,y)$, as shown in FIG.\ref{current}.
For more details about the lattice strain and differential geometry, see, for example, Refs. \cite{Edelen, T.Hughes2013, Kleinert}.



Now, using the linear response theory with the Cartan formalism, we investigate the TCME of WSMs due to dislocation.
We calculate the current density in the presence of TMF and magnetic field up to the linear order.
We use the model of a pair of Weyl fermions with the opposite chirality,
whose Weyl points are at ${\bm k}={\bm \lam}^L$ and ${\bm \lam}^R$ in the momentum space, and
Fermi energies are given by $E=v_{F}\lam^L_0$ and $v_{F}\lam_0^R$, respectively.
Therefore the $4\times{4}$ Hamiltonian is given by
\begin{align}
H({\bm k}):= 
\left(
\begin{array}{cc}
H_{L}({\bm{k}})&0\\
0&H_{R}({\bm{k}})
\end{array}
\right)
\end{align}
with $H_{s}({\bm{k}}):=v_F\left[\chi_{s}({\bm{k}}-{\bm\lam}^{s})\cdot{\bm\sig}-\lam^{s}_0\right]$, 
where $s=L\ {\rm or}\ R$ is the index of the chirality and $\chi_{L(R)}=+1(-1)$,
and $\sigma^i$ is the Pauli matrix.
Well, we calculate the current density in the presence of the external fields.
The calculation is performed by the variation of the effective action, $S_{{\rm{eff}}}[A_i,e^a_i]$,
with respect to the gauge field, as $j^a({\bm r}):= -(e_i^a({\bm r})/|e({\bm r})|)(\delta S_{{\rm eff}}/\delta A_i({\bm r}))$.
The effective action is defined as
\begin{align}
& e^{- S_{{\rm eff}}[A_i ,e^a_i]} := \int \mathcal{D} \psi \mathcal{D}\psi^{\dag}  \exp \left( - S[\psi, \psi^{\dag}, A_i,e^a_{i}] \right),  \nonumber \\
& S[\psi, \psi^{\dag}, A_i,e^a_i] :=\frac{1}{2} \int d\tau d^3r  \left[   \psi^{\dag} (\tau , {\bm r} )\hat{ \mathcal{L} } \psi (\tau , {\bm r} )  + c.c.   \right] , \nonumber \\
& \hat{\mathcal{L}} :=  |e({\bm r})| [ \part / \part \tau - 
H(-\mathrm{i} \nabla_a)], 
\label{eq:eff-action}
\end{align}
where $\psi$ is the fermionic field, $\tau$ and ${\bm r}$ denote the imaginary time and spatial coordinate, respectively,
and $A_i$ 
is the vector potential. 
Here $c.c.$ represents the complex conjugate combined with the change of the sign of the derivative operator $\part/\part \tau$.
Also, in Eq.(\ref{eq:eff-action}),  the Jacobian is given by $|e({\bm r})|:=\det e^a_i({\bm r})$,
and the covariant derivative is $-\mathrm{i}\nabla_a:=e^j_{a}({\bm{r}})(-\mathrm{i}\part_j-e A_j({\bm{r}}))$ with $a=\bar{i}$.
Using Eq.(\ref{eq:eff-action}),
we obtain that the
current density up to the first order of the magnetic field and the TMF is given by
\begin{align}
{\bm j} ({\bm r}) =&\left[  \frac{e^2 v_F    (\lam_{{0}}^R  - \lam_{{0}}^L)}{4 \pi^2}{\bm B}  + \frac{e  v_F   (\lam^R_a - \lam^L_a ) \Lam   }{ 4 \pi ^2}  {\bm T^a} \right], \label{WSM_current} \nonumber \\
\end{align}
at zero temperature and up to the linear order in $\lam_{\mu}^{L(R)}$,
where the details of the calculations are described in Ref.\cite{supp}.
Here, 
The vector representation of the TMF, ${\bm T}^a$, is defined by $({\bm T^a})_i:=(1/2)\vareps^{ijk}T^{a}_{jk}$. 
For the derivation of Eq.(\ref{WSM_current}), we introduced a momentum cutoff scheme
$|{\bm k}-{\bm \lam}^{s}|<\Lam$ for the Weyl node of the chirality $s$.
Physically, ${\Lam}$ corresponds to the momentum range from the Weyl points in which the cone structures of the band of the lattice system is approved.


The first term represents the CME in the presence of the chiral chemical potential (i.e. $\lam^L_0\neq\lam^R_0$),
and then reproduces the previous result for the CME \cite{SonYamamotoPRL}.
On the other hand, the second term in Eq.(\ref{WSM_current}) 
is a new one, which raises 
the TCME;
i.e. the current is generated by the TMF
for the pair of Weyl points which are shifted in the momentum space due to broken time-reversal symmetry (TRS).
This point is in sharp contrast to the usual CME, which requires breaking inversion symmetry.

We comment on the relation between the TCME and the chiral anomaly.
One may expect that when, $\lam^R_{\mu}=-\lam^L_{\mu}$, the TCME is described by the topological $\theta$-term, which is the consequence of the chiral anomaly like the CME and anomalous Hall effect.
However, there is no mixed chiral anomaly term of $U(1)$ field strength and torsion in four-dimensional spacetime
\cite{munoz2010, Parrikar2014}. 
This point is resolved by the observation that the Lorentz symmetry, which is postulated in the calculation scheme in Refs. \cite{munoz2010, Parrikar2014},
is broken in the cutoff scheme used for the derivation of 
the second term of Eq.(\ref{WSM_current}), 
which is correctly applicable to
realistic condensed matter systems.

Now we discuss the consequences and physical pictures of the TCME.
The TCME is realized in two types of lattice dislocations.
(a) {\it case of edge dislocation}: 
$j^x\propto  \Delta \lam_z\ T^z_x$, and 
(b) {\it case of screw dislocation}: $j^z \propto \Delta \lam_z T^z_z$,
with $\Delta \lam_a := \lam^L_a - \lam^R_a$.
Their schematic pictures are shown in FIG.\ref{current}.
These responses can be understood with the following semiclassical picture:
{\it Case} (a): Edge dislocation is regarded as the (0,1,0)-``surface"
of the extra lattice plane made up of the blue and green atoms in FIG.\ref{current}-(a),
which harbors a chiral Fermi arc, when two Weyl points are shifted in the $k_z$-direction.
The electrons in the Fermi arc state are the very origin of the current induced by the edge dislocation.
{\it Case} (b): 
There is a chiral Fermi arc mode on the dislocation line.
The electrons in the mode rotate around the screw dislocation line,
and due to the screw dislocation the rotating motion causes the current along the Burgers vector.

The situation is similar to that of
the three-dimensional integer quantum Hall state (3DIQHS) \cite{Kohmoto1992} with 
dislocation
which is the staking of quantum Hall state layers characterized by 
the vector ${\bm G}_c=(2\pi{n_c}/a)\hat{n}$, where $n_c$, $a$, and $\hat{n}$ are the first Chern number,  
the lattice constant, and the unit vector along the staking direction.
In the 3DIQHS, there are one dimensional $n$ chiral modes along the dislocation line,
when the topological number, $n={\bm{b}}_g\cdot{}{\bm{G}_c}/2\pi$, is nonzero 
\cite{Teo-Kane2010}.
This condition for the chiral modes 
is similar to that for the TCME, ${\bm b}_g\cdot ({\bm \lambda}^L -{\bm \lambda}^R)\neq 0$.

\begin{figure}
 \begin{center}
 \includegraphics[width=80mm]{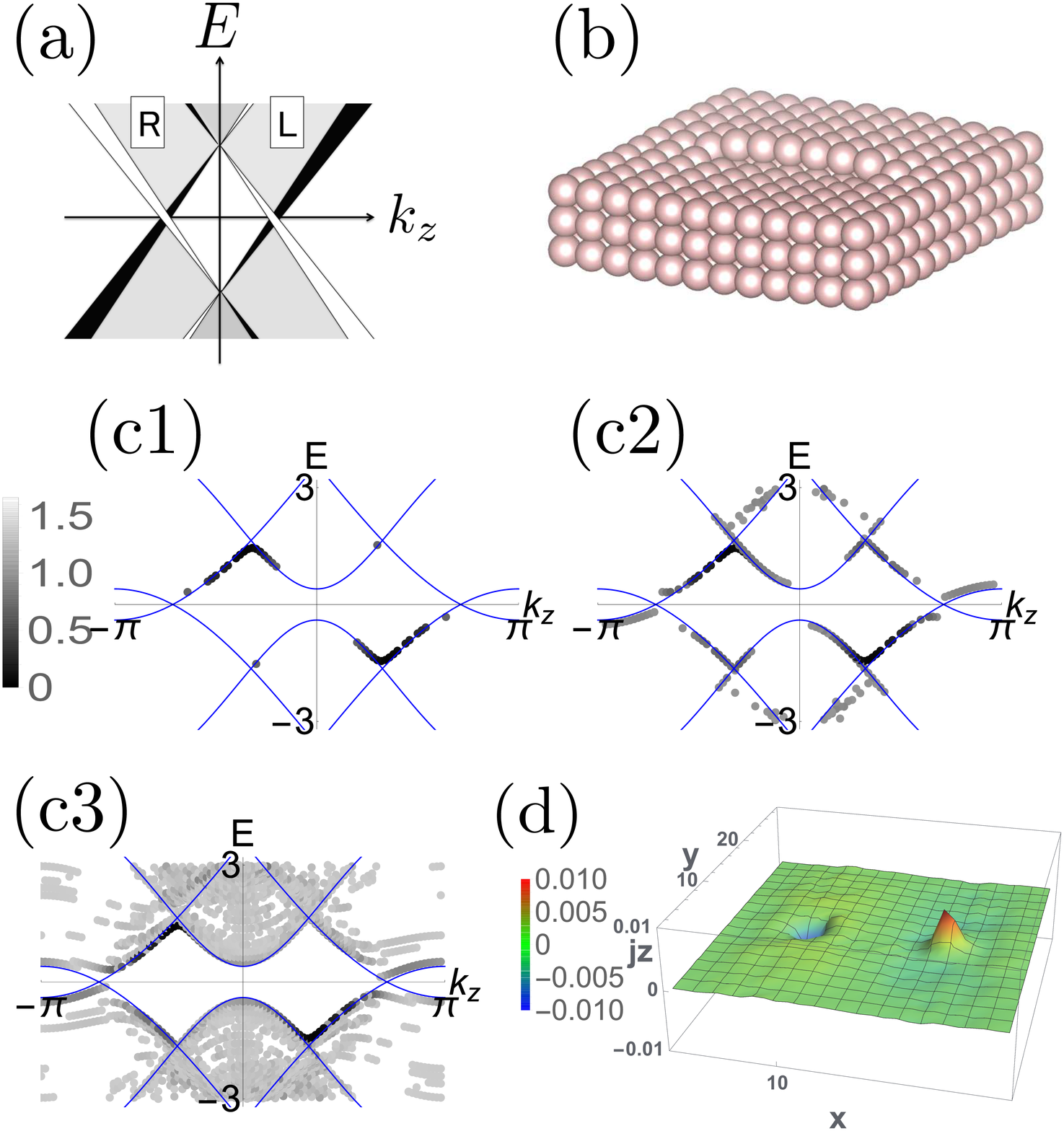}
 \end{center}
\caption{(Color Online)
(a) Schematic picture of the spectrum of the WSM with the screw dislocation.
The black (white) bands
along $E=\pm{m^s_{k_z}}$ represent the relatively higher (lower) density-of-state compared with that of the opposite energy $E=\mp{m^s_{k_z}}$.
(b) Lattice with a pair of screw dislocations with opposite Burgers vectors.
(c)
Numerical result for the spectrum of the WSMs with a pair of screw dislocations.
The blue curves are the envelope of the bulk spectrum.
The opacity of the dots 
represents the expectation value of $|{\bm{\rho}}-{{\bm{l}}^{dis}}|^2/L_xL_y$ (see the gray scale bar).
In the figures (c1-3), the modes with this value smaller than $0.1$, $0.15$, and $0.2$ are plotted.
(d)
Current density along $z$-direction, $j_z(x,y)$.
}
 \label{spectrum}
\end{figure}

However, there are the following significant differences.
The chiral modes of the 3DIQHS are exponentially localized at the dislocation, and
separated from the bulk higher-energy states, while,
 as will be shown below, the chiral modes of the WSMs 
exhibit power-law decay.
We call them quasi-localized modes.
Moreover their spectrum is not isolated from the bulk spectrum but appears as its envelope as shown in FIG.\ref{spectrum}a,
and therefore they 
can be easily mixed with the bulk modes.

{\it Spectral asymmetry and ground state current---}
We confirm the TCME due to dislocation by using an alternative approach other than the linear response theory based on (\ref{WSM_current}).
Our approach here is to calculate explicitly the spectrum and the eigenstates of the Weyl Hamiltonian with dislocation and the ground state current.
We also show that the quasi-localized modes along the envelope of the bulk spectrum contribute to the effect.
For simplicity, we set $v_F=e=1$ and assume 
$\lam^L_0=\lam^R_0=0$,
and the Weyl points lie symmetrically on the $k_z$-axis, ${\bm \lam}^L=-{\bm \lam}^R=\lam\hat{z}$.
In the presence of the screw dislocation at $x=y=0$ along $z$-axis, of which Burgers vector is ${\bm b}_g=-b_g\hat{z}$, the vielbeins are given by 
$e^z_{\bar{x}}=-b_gy/ 2 \pi \rho^2,\ e^z_{\bar{y}}=b_gx/2\pi \rho^2$, and $e^{\mu}_{a}=\delta^{\mu}_a$ for others,
with $\rho=\sqrt{x^2+y^2}$ \cite{Edelen}.
Even with the dislocation, $k_z$ remains a good quantum number.
Then, when $k_z$ is fixed,
the Hamiltonian is equivalent to
that of two dimensional massive Dirac model in the presence of the magnetic flux at the origin, whose amplitude is $\Phi_{k_z}=k_zb_g$,
\begin{align}
&H^{{\rm screw}}_{s,k_z}=\chi_{s} \left[  H^{\perp}_{k_z}  + m^{s}_{k_z} \sig^z\right], \nonumber \\
&H^{\perp}_{k_z}= \left(-\mathrm{i}\part_x- \frac{\Phi_{k_z} y}{2\pi \rho^2} \right)\sig^x + \left(-\mathrm{i}\part_y+ \frac{\Phi_{k_z}x}{2\pi \rho^2}\right)\sig^y,
\end{align}
with the mass $m^{s}_{k_z}=k_z-\chi_{s}\lam$.
The equivalence of a screw dislocation and momentum-dependent magnetic field has also been pointed out in Refs. \cite{Imura2011, Parrikar2014}.

The spectrum of $H^{{\rm screw}}_{s,k_z}$ consists of two types of eigenstates:
one with the eigenenergies satisfying $|E|>|m^{s}_{k_z}|$ and the other one with $E=\pm{m^{s}_{k_z}}$.
The former does not contribute to the ground state current owing to the one-to-one correspondence between $E^+>|m^{s}_{k_z}|$ and $E^-<-|m^{s}_{k_z}|$ modes as $E^+=-E^-$,
and between the states of Weyl nodes with the opposite chiralities \cite{supp}.
On the other hand, the latter does contribute owing to asymmetry, i.e. the absence of one-to-one correspondence between $E={m^{s}_{k_z}}$ and $E=-{m^{s}_{k_z}}$ modes.
This asymmetry is called the parity anomaly \cite{Jackiw1984, Haldane1988}.
The asymmetric spectrum consists of discrete modes whose wavefunctions exhibit power-law decay, and
continuum scattering modes which spread over the whole system \cite{Kiskis1977,Aharonov1979,supp}.
The schematic picture of the density-of-state of the full spectrum is shown in FIG.\ref{spectrum}a.
Moreover, the ground state current calculated from the asymmetric spectrum is $J^z=-L_zb_g\Lam\lam/2\pi^2$, which coincides with the expression obtained directly from the linear response theory (\ref{WSM_current})
\cite{supp}.

%

{\it Numerical calculation ---}
We confirm the spectrum asymmetry and
the TCME for realistic lattice models by numerical calculations.
We use the tight-binding model of WSMs \cite{Vazifeh2013} generalized to the case with dislocation,
\begin{align}
&H =
\sum_{{\bm r}}  \left[   {\mathrm i}t\sum_{i=x,y,z} c^{\dag}_{{\bm r}+\hat{i}+\delta_{i,z}b_g\Theta({\bm r})}\Gamma^i c_{{\bm r}}  +r \left( 3c^{\dag}_{{\bm r}}\Gamma^4c_{\bm r}\phantom{\sum_{i=x,y,z}} \right. \right. \nonumber\\
& \left. \left. -  \sum_{i=x,y,z} c^{\dag}_{{\bm r}+\hat{i}+\delta_{i,z}b_g\Theta({\bm r})}\Gamma^4 c_{{\bm r}} \right)
+ \frac{d}{2} c^{\dag}_{{\bm r}}\Gamma^{12} c_{{\bm r}} \right]+h.c., \label{numerical_hamil}
\end{align}
where the $4\times{4}$-matrices, $\Gamma^i$, satisfy the $SO(5)$ Clifford algebra $\{\Gamma^i,\Gamma^j\}=2\delta_{ij}$ \cite{Murakami2004},
$\Gamma^{ij}:=[\Gamma^i,\Gamma^j]/2i$,
${\bm r}=(x,y,z)$ and $\hat{i}$ denote the position of the atoms and the $x^i$-direction unit vector, respectively,
and 
$t,r$, and $d$ are the real parameters,
and we suppose the lattice constant as $1$ and lattice size $L_x\times{L_y}\times{L_z}$.
We introduced a pair of screw dislocations along $z$-direction with opposite Burgers vector at $\pm{{\bm{l}}^{dis}}=\pm({l^{dis}_x},0)$
as shown in FIG.\ref{spectrum}-b, by
sliding the hopping directions in the first and third terms of Eq. (\ref{numerical_hamil})
as $\Theta({\bm r})=-1$ for the region $x=0$, $-l^{dis}_x<y<l^{dis}_x$, while $\Theta({\bm r})=0$ for other regions.
We numerically diagonalized this model and obtained the spectrums and current.
Here the material parameters are set as $t=r=1$ and $d=3.6$.
The lattice constant is $1$ and the amplitudes of the Burgers vectors is set as $b_g=1$.
For the calculation, we imposed the open boundary condition along the $x-$ and $y-$directions
and periodic boundary condition along the $z-$direction,
and set $L_x=L_y=4l^{dis}_x=38$ and $L_z=100$.

As shown in FIG.\ref{spectrum}c1-3, we obtained the asymmetric spectrum in agreement with the analytic calculation.
The asymmetric modes are localized at the dislocation line.
The quasi-localized chiral modes are not isolated from the bulk but easily mixed with the bulk modes  (FIG.\ref{spectrum}c1-3). 
The current density at zero temperature is shown in FIG.\ref{spectrum}d.
We obtained the upward current along the screw dislocation and downward current along the anti-screw dislocation
due to the TCME.
The total current per the unit length toward $z$-direction due to one dislocation line is $J^z/L_z=0.087$,
which is calculated by the summation of the current density in the $x>0$ half-plane,
and this value is in the same order as that estimated from the linear response theory (\ref{WSM_current}), $J^z/L_z\sim0.1$.
For the estimation, we set the cutoff as $\Lam\sim1/{\rm{(lattice\ constant)}}=1$.


{\it No-go theorem of CME ---}
The existence of the TCME in the realistic lattice system may seem to contradict with the no-go theorems of the ground state current \cite{Vazifeh2013, Yamamoto_2015}.
However, they prohibit the total current, but not the local current density. 
Therefore, the current along 
the dislocation line can exist, as we found\cite{supp}.

{\it Experimental implication ---}
Here we present two experimental setups to observe the TCME in TRS-broken WSMs, for which Eu$_2$Ir$_2$O$_7$ \cite{Sushkov_EIO2015} and YbMnBi$_2$\cite{Borisenko} are candidate materials.
The first one is a scanning SQUID measurement, 
which can detect weak inhomogeneous magnetic fields \cite{Vasyukov2013, Wang948}. 
If there is a pair of dislocation, the circulating current occurs.
The magnitudes of the current and the induced magnetic field are estimated as $I\sim10^{-5}{\rm{A}}$ and $B\sim10^{-7}{\rm{T}}$, respectively,
for both Eu$_2$Ir$_2$O$_7$ and YbMnBi$_2$.
Here we used Eq.(\ref{WSM_current}) and the material parameters, $v_F\sim10^5m/s$ and $a\sim10{\rm{\AA}}$
and set $b_g=a$ and $\lam\sim\Lam\sim1/a$, where $a$ is the lattice constant.
Also, for the estimation of the magnitude of the magnetic field, we used 
a typical value of inter-distance between dislocations,  $10^{5}\mathrm{\AA}$ \cite{dis3}.
It is feasible to detect $B\sim10^{-7}{\rm{T}}$  via the scanning SQUID.

\begin{figure}
 \begin{center}
 \includegraphics[width=70mm]{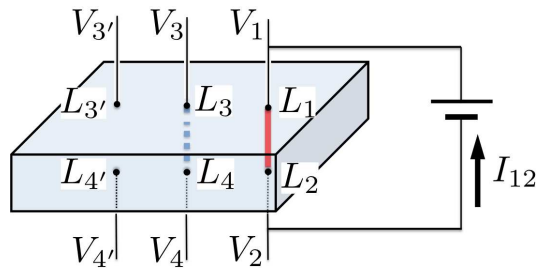}
 \end{center}
\caption{(Color Online)
Experimental setup for the nonlocal transport due to the TCME.
The thick red solid (blue dashed) line represents the (anti-)dislocation line.
The leads are attached at the black points, $L_i$($i=1,2,3,4,3',4'$).
The line $L_1L_2$, $L_3L_4$, and $L_{3'}L_{4'}$ are parallel and have the same length.
Here we suppose $L_1L_3<L_1L_{3'}$.
$V_i$ is the voltage at $L_i$, and $I_{12}$ is the current.
}
 \label{fig_exp}
\end{figure}

The second one is a nonlocal transport phenomenon, which was observed in quantum Hall materials \cite{McEuen1990, Wang1991}. 
The experimental setup is shown in FIG.\ref{fig_exp}. 
If the bulk contributions are completely negligible and there are only the chiral modes at the dislocation lines,
$V_{34}:=V_3-V_4=0$ despite $I_{12}\neq0$, then, the nonlocal resistivity $R_{12,34}:=V_{34}/I_{12}$ is equal to zero \cite{McEuen1990}. 
On the other hand, if 
the nonlocal transport is negligible, $V_{34}>V_{3'4'}$ holds for $L_1L_3<L_{1}L_{3'}$ when $I_{12}\neq0$.
Therefore, if $R_{12,34}<R_{12,3'4'}$ is observed, it is the fingerprint of the chiral current due to the TCME.
The effect can be discriminated from any previously reported conventional transport induced by dislocation\cite{dis0,dis1,dis2,dis3,dis4,dis5,dis6,dis7,dis8,dis9}.

We also comment on effects of impurities and disorientation of the dislocation. 
First, the current due to the TCME is expected to be robust against weak disorder. 
It is because that Eq.(\ref{WSM_current}) is independent of the scattering time,
like the intrinsic contribution to the anomalous Hall effect \cite{Nagaosa_AHE}.
More precisely, the current 
is  due to the edge modes in the Fermi arc
 on the surface of WSMs,
and these modes are supported by the Weyl points, 
which are protected by the Chern number, and hence, robust againt weak disorder.



Next, in real experimental setups, it is difficult to align the dislocation line orthogonal (parallel) to the line connecting the Weyl nodes exactly in the case (a) (case (b)).
Even when they are not orthogonal (parallel), as long as they are not parallel (orthogonal), the current parallel to the dislocation line still exists.
Supposing that the dislocation line is parallel to the $z$-axis,
the current is given by
$J^z=ev_FL_z  \Lambda ({\bm \lambda}^R-{\bm \lambda}^L)\cdot {\bm b_g}/4\pi^2$ in the both cases (a) and (b).

{\it Summary ---}
In this letter, we have discussed the TCME in WSMs caused by dislocation.
We have confirmed that it is possible to occur and experimentally observable in realistic materials, 
and argued that the Lorentz symmetry breaking is important for it. 

\begin{acknowledgments}
{\it Acknowledgement ---}
We thank H. Fujita, Y. Kikuchi, T. Kimura,  K. Ohmori, D. Schmeltzer, K. Shiozaki,
and A. Shitade 
for fruitful discussions. 
We also thank one of the referees for suggesting scanning SQUID measurements.
This work is supported by the Grant-in-Aids for Scientific
Research from MEXT of Japan [Grant No. 23540406, No. 25220711, and No. 25103714 (KAKENHI on Innovative Areas “Topological Quantum Phenomena"), No. 15H05852 (KAKENHI on Innovative Areas  “Topological Materials Science")].
HS is supported by a JSPS Fellowship for Young Scientists (14J00647).
\end{acknowledgments}

\phantom{*}

\begin{center}
\Large{Supplemental Material}
\end{center}

\appendix
\setcounter{equation}{0}
\renewcommand{\theequation}{S-\arabic{equation}}

\section{Derivation of Eq.(3)}
In this section, we derive 
the expression for the current in the presence of
the magnetic and torsional magnetic responses of the current density, Eq.(3).
The derivation consists of two steps:
first we derive the expression for the Green function in the presence of the gauge field and vielbein using the gradient expansion Eq.(\ref{Green_1st}),
and next we calculate the current density by using Eq.(\ref{Green_1st}) and obtain Eq.(\ref{WSM_current_1_final}), which is equivalent to Eq.(3).

First, we calculate the single-electron Green function.
The Green function in the presence of the gauge field and vielbein which is defined by
\begin{align}
&G( \tau_1 , {\bm r}_1 , \tau_2, {\bm r}_2)  \nonumber \\
&:= \frac{ \int \mathcal{D} \psi \mathcal{D}\psi^{\dag}   \psi(\tau_1 , {\bm r}_1 ) \psi^{\dag} (\tau_2 , {\bm r}_2 ) \exp \left( - S[\psi, \psi^{\dag}, A_i,e^a_i] \right) }{ \int \mathcal{D} \psi \mathcal{D}\psi^{\dag}  \exp \left( - S[\psi, \psi^{\dag}, A_i,e^a_i] \right) }.
\end{align}
Then, the following differential equation holds :
\begin{align}
&\frac{1}{2} \left[   \hat{\mathcal{L}}(\vareps_N,{\bm r}_1 , -\mathrm{i} \part_{{\bm r}_1} ) G(\vareps_N, {\bm r}_1,{\bm r}_2 )  \right. \nonumber \\
&+\left. G(\vareps_N, {\bm r}_1,{\bm r}_2 )   \overleftarrow{\hat{\mathcal{L}}}^* ( - \vareps_N, {\bm r}_2, -\mathrm{i} \part_{{\bm r}_2})  \right] =\delta^{(3)}({\bm r}_1 - {\bm r}_2) ,  \label{dyson}
\end{align}
with $ \hat{\mathcal{L}}(\vareps_N, {\bm r}, -\mathrm{i} \part_{\bm r}  ) :=  |e({\bm r})| [ \mathrm{i} \vareps_N - H(-\mathrm{i} \nabla_a)]  -e A_0]$, and $|e({\bm r})|:=\det e^a_i({\bm r})$
Here $\vareps_N=(2N+1)\pi T$ is the Fermionic Matsubara frequency with the temperature $T$,
and $G(\vareps_N, {\bm r}_1,{\bm r}_2 ):=\int_0^{\beta}G(\tau, {\bm r}_1, 0, {\bm r}_2 ) e^{-\mathrm{i} \vareps_N \tau}d\tau$ is the Fourier component of the Green function.
Now, using the spatial Wigner transformation defined as $\tilde{f}({\bm R},{\bm p}) := \int d^3{\bm r} e^{-\mathrm{i}{\bm r} \cdot {\bm p}} f({\bm R} + {\bm r}/2, {\bm R} - {\bm r} /2)$,
Eq. (\ref{dyson}) is rewritten into
\begin{align}
 &\frac{1}{2} \left[ \hat{\mathcal{L}}(\vareps_N , {\bm R}, {\bm p}) e^{\frac{\mathrm{i}}{2} (\overleftarrow{\partial_{\bm R}} \overrightarrow{\partial_{\bm p}} -\overleftarrow{\partial_{\bm p}} \overrightarrow{\partial_{\bm R}}) }   \tilde{G}(\vareps_N , {\bm R}, {\bm p})  \right.\nonumber \\
 &\left. +    \tilde{G}(\vareps_N , {\bm R}, {\bm p})  e^{\frac{\mathrm{i}}{2} (\overleftarrow{\partial_{\bm R}} \overrightarrow{\partial_{\bm p}} -\overleftarrow{\partial_{\bm p}} \overrightarrow{\partial_{\bm R}}) }   \hat{\mathcal{L}}(\vareps_N , {\bm R}, {\bm p}) \right]
  = 1 .
\end{align}
In the gradient expansion up to the first order in $\part_i A_j$ or $\part_i e^a_j$, the Green function becomes
\begin{align}
&\tilde{G}(\vareps_N , {\bm R}, {\bm p}) = \tilde{G}^{(0)}(\vareps_N , {\bm R}, {\bm p})  + \tilde{G}^{(1)}(\vareps_N , {\bm R}, {\bm p}) + \cdots, \nonumber \\
&\tilde{G}^{(0)}(\vareps_N , {\bm R}, {\bm p}) = \frac{1}{|e({\bm R})|} \left[ \mathcal{L}_0 (\vareps_N , {\bm \pi}) \right]^{-1}_{\pi_a = e^i_a({\bm R}) (p_i - e A_i({\bm R}))}, \nonumber \\
& \tilde{G}^{(1)}(\vareps_N , {\bm R}, {\bm p}) \nonumber \\
&=  \frac{\mathrm{i}}{2|e({\bm R})|}  \mathcal{L}^{-1}_0 (\vareps_N , {\bm \pi}) \frac{ \part \mathcal{L}_0 (\vareps_N , {\bm \pi})}{\part \pi_a } \mathcal{L}^{-1}_0 (\vareps_N , {\bm \pi})  \frac{ \part \mathcal{L}_0 (\vareps_N , {\bm \pi})}{\part \pi_b} \nonumber \\
&\cdot \left. \mathcal{L}^{-1}_0 (\vareps_N , {\bm \pi})  \left[ e F_{ab}({\bm R}) + T^c_{ab}({\bm R}) \pi_c \right] \right|_{\pi_a = e^i_a({\bm R}) (p_i - e A_i({\bm R}))}. \label{Green_1st}
\end{align}
Here $\pi_a := e^i_a({\bm R}) (p_i - e A_i({\bm R}))$ is the gauge-invariant mechanical momentum, while $p_i$ is the canonical momentum, and the field strength and the torsion with the indices of the local orthogonal coordinate are, respectively, defined as
$F_{ab}:=e^i_a e^j_b F_{ij}$ and $T^c_{ab}:=e^i_a e^j_b T^c_{ij}$.
The free Lagrangian is defined as $ \mathcal{L}_0 (\vareps_N , {\bm \pi}) := \mathrm{i} \vareps_N - H({\bm \pi})$. 

Next, 
using Eq. (\ref{Green_1st}), we calculate the current density and derive Eq.(2).
The current density are defined by
$j^a({\bm r}):= -(e_i^a({\bm r}))/|e({\bm r}))|) (\delta S_{{\rm eff}} /\delta A_i({\bm r}))$.
Therefore 
\begin{align}
&j^{\bar{1}}({\bm R})\nonumber \\
&=\frac{e^{\bar{1}}_{i}}{|e({\bm R})|}\left. \frac{ \int \mathcal{D} \psi \mathcal{D}\psi^{\dag}   \psi^{\dag}(\tau, {\bm r}_1 ) \frac{1}{2} \frac{\delta {\hat {\mathcal L}}}{\delta A_i} \psi (\tau, {\bm r}_2 )  e^{ - S }} { \int \mathcal{D} \psi \mathcal{D}\psi^{\dag}  e^{ - S }} \right|_{{\bm r}_1,{\bm r}_2 \to {\bm R},}+ c.c. \nonumber \\
&=\frac{eT}{2} \sum_{N} {\rm Tr} \left[ \left. \frac{\part \mathcal{L}_0(\epsilon_N,{\bm \pi})}{\part \pi_{\bar{1}}}\right|_{\pi_a=e^i_a({\bm r}_2)(-\mathrm{i}\part_{r_2^i}-eA_i({\bm r}_2))} \right. \nonumber \\
&\left. \left. G(\vareps_N, {\bm r}_2,{\bm r}_1) \right] \right|_{{\bm r}_1,{\bm r}_2 \to {\bm R}}+ c.c. \nonumber \\
&=\frac{eT}{2} \sum_{N}\int \frac{d^3p}{(2\pi)^3} {\rm Tr} \left[ \frac{\part \mathcal{L}_0(\epsilon_N,{\bm \pi})}{\part \pi_{\bar{1}}}\right|_{\pi_a=e^i_a({\bm R})(p_i-eA_i({\bm R}))}
\nonumber \\
& \left. e^{\frac{\mathrm{i}}{2} (\overleftarrow{\partial_{\bm R}} \overrightarrow{\partial_{\bm p}} -\overleftarrow{\partial_{\bm p}} \overrightarrow{\partial_{\bm R}}) }   \tilde{G}(\vareps_N , {\bm R}, {\bm p})
\right] + c.c.
\end{align}
Here ${\rm Tr}$ means the trace over the band indices, we used 
that $\delta \hat {\mathcal L}/\delta A_i=e|e({\bm r})|e^i_a({\bm r}) \part L_0/\part \pi_a$ and the second line does not depend on $\tau$ due to
imaginary time-translation symmetry.
Note that ${\bm \pi}$ of the third line is the operator though that of the fourth one is the c-number.
Using Eq.(\ref{Green_1st}), up to the first order in $\part_i A_j$ or $\part_i e^a_j$, the expression of the current density becomes
\begin{align}
&j^{\bar{1}}({\bm R}) = j^{\bar{1}(0)}({\bm R})+j^{\bar{1}(1)}({\bm R}),
\end{align}
with the zeroth-order terms,
\begin{align}
&j^{\bar{1}(0)}({\bm R})\nonumber \\
&=\frac{eT}{2}\sum_{N}\int \frac{d^3p}{(2\pi)^3}{\rm Tr}\left[\left. \frac{\part \mathcal{L}_0(\epsilon_N,{\bm \pi})}{\part\pi_{\bar{1}}} \right|_{\pi_a=e^i_a({\bm R})(p_i-eA_i({\bm R}))}  \right. \nonumber \\
& \left. \tilde{G}^{(0)}(\vareps_N , {\bm R}, {\bm p}) \right] +c.c. \label{current_0th}
\end{align}
and the first-order terms,
\begin{align}
&j^{\bar{1}(1)}({\bm R}) \nonumber \\
&=\frac{eT}{2}\sum_{N}\int \frac{d^3p}{(2\pi)^3}{\rm Tr}\left[\left. \frac{\part \mathcal{L}_0(\epsilon_N,{\bm \pi})}{\part\pi_{\bar{1}}} \right|_{\pi_a=e^i_a({\bm R})(p_i-eA_i({\bm R}))}  \right. \nonumber \\
& \left. \tilde{G}^{(1)}(\vareps_N , {\bm R}, {\bm p}) \right] \nonumber \\
& + \frac{\mathrm{i}eT}{4}\sum_{N}{\rm Tr}\left[ \left. \frac{\part \mathcal{L}_0(\epsilon_N,{\bm \pi})}{\part\pi_{\bar{1}}} \right|_{\pi_a=e^i_a({\bm R})(p_i-eA_i({\bm R}))} \right. \nonumber \\
& \left.  (\overleftarrow{\partial_{\bm R}} \overrightarrow{\partial_{\bm p}} -\overleftarrow{\partial_{\bm p}} \overrightarrow{\partial_{\bm R}})  \tilde{G}^{(0)}(\vareps_N , {\bm R}, {\bm p}) \right] +c.c.. \label{current_1st}
\end{align}
The zeroth-order terms (\ref{current_0th}) can be rewritten as
\begin{align}
j^{\bar{1}(0)}({\bm R})=e\sum_n\int\frac{d^3\pi}{(2\pi)^3}  v^{n,\bar{1}}({\bm \pi})  n_F(\vareps_{n, {\bm \pi}}). \label{j_1_0_final}
\end{align}
For the derivation,
we inserted the identity, $1_{{\bm \pi}}=\sum_n \ket{u^n_{{\bm \pi}}}\bra{u^n_{{\bm \pi}}}$,
between $\part \mathcal{L}_0/\part \pi_{\bar{1}}$ and $\tilde{G}^{(0)}$ in Eq.(\ref{current_0th}),
and used the formula, 
$\sum^{\infty}_{N=-\infty}\left[ 1/(\mathrm{i}\vareps_N-t)+1/(-\mathrm{i}\vareps_N-t) \right]=(1-2n_F(t))/T$,
for the summation over the Matsubara frequency,
and$ \int d^3p=|e({\bm R})|\int d^3\pi$.
Here, $n$ is the band index,
$\vareps_{n,{\bm \pi}}$ is the energy, $v^{n,a}({\bm \pi}):=\part \vareps_{n,{\bm \pi}}/\part \pi_a$ is the group velocity,
$n_F(\vareps):=1/(e^{\vareps/T}+1)$ is the Fermi distribution function,
and $\ket{u^n_{{\bm \pi}}}$ is the Bloch state.
This term corresponds to the summation of all contributions to the current from the electrons in the occupied states in the absence of magnetic and torsional magnetic field.

Now, we move on the calculation of Eq.(\ref{current_1st}).
The sum of the second term of Eq.(\ref{current_1st}) and its complex conjugate is zero,
since 
$\left[ \cdots \right]^*=\left. \left[ \cdots \right] \right|_{\vareps_N\to-\vareps_N}$.
Then, by using Eq.(\ref{Green_1st}), we obtain
\begin{align}
&j^{\bar{1}(1)}({\bm R}) \nonumber \\
&=\frac{\mathrm{i}eT}{4}\sum_{N}\int \frac{d^3{\pi}}{(2\pi)^3} {\rm Tr}\left[
\frac{ \part \mathcal{L}_0 (\vareps_N , {\bm \pi})}{\part \pi_{\bar{1}} } \mathcal{L}^{-1}_0 (\vareps_N , {\bm \pi}) \frac{ \part \mathcal{L}_0 (\vareps_N , {\bm \pi})}{\part \pi_a } \right. \nonumber \\
&\left. \mathcal{L}^{-1}_0 (\vareps_N , {\bm \pi})  \frac{ \part \mathcal{L}_0 (\vareps_N , {\bm \pi})}{\part \pi_b} \mathcal{L}^{-1}_0 (\vareps_N , {\bm \pi}) \right] \left[ e F_{ab}({\bm R}) + T^c_{ab}({\bm R}) \pi_c \right] \nonumber \\
&+c.c..
\end{align}
Moreover, inserting the identities, $1_{{\bm \pi}}=\sum_n \ket{u^n_{{\bm \pi}}}\bra{u^n_{{\bm \pi}}}$,
we obtain
\begin{align}
&j^{\bar{1}(1)}({\bm R}) \nonumber \\
&=\frac{-\mathrm{i}eT}{4}\sum_{N,n,m,l}\int \frac{d^3{\pi}}{(2\pi)^3} 
\Braket{n|\frac{\part H}{\part \pi_{\bar{1}}}|m}
\Braket{m|\frac{\part H}{\part \pi_{a}}|l} \nonumber \\
&\times\Braket{l|\frac{\part H}{\part \pi_{b}}|n}
\frac{1}{(\mathrm{i}\vareps_N-\vareps_n)(\mathrm{i}\vareps_N-\vareps_m)(\mathrm{i}\vareps_N-\vareps_l)} \nonumber \\
&\times \left[ e F_{ab}({\bm R}) + T^c_{ab}({\bm R}) \pi_c \right] + c.c.,
\end{align}
where the indices ${\bm \pi}$ are omitted like $\vareps_n:=\vareps_{n,{\bm \pi}}$ and $\ket{n}:=\ket{u^n_{{\bm \pi}}}$.
There are three types of contributions to the summation over the band indices $n,m,l$:
(a) all the three are the same, 
(b) two of them are the same and the other is different,
and (c) each one is different respectively. 
However the contribution (a) is found to be zero because of the antisymmetry of $\left[ e F_{ab}({\bm R}) + T^c_{ab}({\bm R}) \pi_c \right]$ under $a\leftrightarrow b$.
Moreover, the contribution (c) is also zero, since 
our model of the WSM, Eq.(1),
consists of two two-band Hamiltonians independent of each other,
and then the overlap of
three or more bands is zero. 
Therefore, we have only to consider the contribution (b), and then obtain
\begin{align}
&j^{\bar{1}(1)}({\bm R}) \nonumber \\
&= \frac{-\mathrm{i}e}{4}\sum_{n} \int \frac{d^3{\pi}}{(2\pi)^3} \left[ e F_{ab} + T^c_{ab} \pi_c \right] (M_{\bar{1}ab}+M_{ab\bar{1}}+M_{b\bar{1}a})\nonumber\\
& + c.c.,\label{current_M}
\end{align}
with
\begin{align}
M_{abc}
&:= v^n_{a}n'_F(\vareps_n) \Braket{n,b|(\vareps_n-H)|n,c} \nonumber \\
& +v^n_{a}\Braket{n,b|n_F(H)|n,c}-v^n_{a}n_F(\vareps_n)\Braket{n,b|n,c} , \label{M_def}
\end{align}
where we used the abridged notation,
$\ket{n,a}:=\Ket{\frac{\part u^n_{\bm \pi}}{\part \pi_{a}}}$.
For the derivation of Eqs.(\ref{current_M},\ref{M_def})
we used the formulae $\sum_{N=-\infty}^{\infty}\frac{1}{( \mathrm{i}\vareps_N-t)^2(\mathrm{i}\vareps_N-s)}=\frac{tn'_F(s)-sn'_F(t)+n_F(s)-n_F(t)}{T(t-s)^2}$,
$\Braket{n|\part H/\part \pi_a|m}=(\vareps^m-\vareps^n)\Braket{n|m,a}$ for $n\neq m$,
and $\sum_mf(\vareps_m)\Braket{n,b|n,c}=\Braket{n,b|f(H)|n,c}$ for any function $f$.
Moreover, using the relationship $(M_{abc})^*=M_{acb}$, we obtain
\begin{align}
&j^{\bar{1}(1)}({\bm R}) \nonumber \\
&= \frac{-\mathrm{i}eT}{2}\sum_{n}\int \frac{d^3{\pi}}{(2\pi)^3}  \left[ e F_{\bar{2}\bar{3}} + T^d_{\bar{2}\bar{3}} \pi_d \right] \vareps^{abc}M_{abc},
\end{align}
where $\vareps^{abc}$ is the antisymmetric symbol.
Furthermore, 
since $v^n_{a}n'_F(\vareps_n)=\part n_F(\vareps_n)/\part \pi_a$
and only antisymmetric parts of $M_{abc}$ contribute,
using integration by parts,
we find
\begin{align}
&j^{\bar{1}(1)}({\bm R}) \nonumber \\
&=\mathrm{i}e\sum_{n}\int \frac{d^3{\pi}}{(2\pi)^3}  \left[ e F_{\bar{2}\bar{3}} + T^d_{\bar{2}\bar{3}} \pi_d \right] \vareps^{abc} v^n_{a}n_F(\vareps_n)\Braket{n,b|n,c} \nonumber \\
&+\frac{\mathrm{i}e}{2}\sum_{n}\int \frac{d^3{\pi}}{(2\pi)^3}  T^a_{\bar{2}\bar{3}}  \vareps^{abc} \vareps_n n_F(\vareps_n)\Braket{n,b|n,c}.
\end{align}
Using  the Berry curvature is defined by
$\Omega^{n}_a := -\mathrm{i} \vareps^{abc} \braket{ n, b | n,c} $,
and the vector representation of the TMF, ${\bm T}^a$, is defined by $ T^a_i:=(1/2)\vareps^{ijk}T^{a}_{jk}$,
it can be rewritten as
\begin{align}
&j^{\bar{1}(1)}({\bm R}) \nonumber \\
&=- e \sum_n  \int \frac{d^3 \pi}{ (2\pi)^3}  \left( {\bm v}^n \cdot  {\bm \Omega}^{ n}\right) (eB_{\bar{1}}+  T^a_{\bar{1}} \pi_a )  n_F(\vareps_{n}) \nonumber \\
&-\frac{e}{2} \sum_n  \int \frac{d^3 \pi}{ (2\pi)^3} \Omega^{n}_a T^a_{\bar{1}}\vareps_nn_F(\vareps_n). \label{BF_current_1}
\end{align}
It is noted that 
the term containing $B_{\bar{1}}$ is equal to
the expression for the CME 
derived by Son and Yamamoto \cite{SonYamamotoPRL},
and the others are new terms that represent the current induced by the torsion.
Neglecting the last term, which is, as we will discuss later, less important than the others in the case of WSMs,
Eq.(\ref{BF_current_1}) can also shortly derived from the
substitution of the magnetic field or the field strength in the absence of the vielbein,
$-\mathrm{i}[(-\mathrm{i}\part_2-eA_2),(-\mathrm{i}\part_3-eA_3)]=eB_{1}$ with 
the field strength in the presence of the vielbein, $-\mathrm{i}[-\mathrm{i}\nabla_{\bar{2}},-\mathrm{i}\nabla_{\bar{3}}]=eB_{\bar{1}}+T^a_{\bar{1}}(-\mathrm{i}\nabla_a)$,
where $[U,V]:=UV-VU$ is the commutator.
This justifies the analogy between the TMF and the magnetic field.

Finally, we substitute the energy, group velocity, and Berry curvature of the model of the WSM, (1), into Eq.(\ref{BF_current_1})
and derive Eq.(3).
We characterize the four bands of the Hamiltonian (1) as $n=(s,\pm)$, with $s=L\ {\rm or}\ R$,
where $s$ is the index of the chirality and $+(-)$ means the higher (lower) band of the Weyl cone.
Then, their energy, group velocity, and Berry curvature are given by
\begin{align}
&\vareps^{s,\pm}({\bm k}) = v_F \left[ \pm   |{\bm k} - {\bm \lam}^{s}| - \lam^{s}_0 \right]  \nonumber \\
&{\bm v}^{s,\pm} ({\bm k}) = \pm \frac{v_F({\bm k} - {\bm \lam}^s)}{|{\bm k} - {\bm \lam}^s|} \nonumber \\
&{\bm \Omega}^{s,\pm} ({\bm k}) = \pm \chi_{s}\frac{{\bm k} - {\bm \lam}^{s}}{{2|{\bm k} - {\bm \lam}^s|^3}}. \label{BC,vel}
\end{align}
Using Eqs.(\ref{BF_current_1}, \ref{BC,vel}),
we obtain
\begin{align}
&j^{\bar{1}(1)} ({\bm R}) \nonumber \\
&=\left[  \frac{e^2 v_F    (\lam_{{0}}^R  - \lam_{{0}}^L)}{4 \pi^2}B_{\bar{1}}({\bm R})  + \frac{e  v_F   (\lam^R_a - \lam^L_a ) \Lam   }{ 4 \pi ^2}  T^a_{\bar{1}}({\bm R})  \right], \label{WSM_current_1_final} \nonumber \\
\end{align}
at zero temperature and up to the linear order in $\lam_{\mu}^{L(R)}$.
For the derivation of Eq.(\ref{WSM_current_1_final}), we introduced a momentum cutoff scheme
$|{\bm k}-{\bm \lam}^{s}|<\Lam$ for the Weyl node of the chirality $s$.
Physically, ${\Lam}$ corresponds to the momentum range from the Weyl points in which the cone structures of the band of the lattice system is approved.
Note that the last term of Eq.(\ref{BF_current_1}) yields second(or more)-order contributions in $\lam_{\mu}^{L(R)}$, and then less important as mentioned before.
Eq.(\ref{WSM_current_1_final}) is the correction of current due to the TMF and magnetic field and is equivalent to Eq.(3),
then the derivation of Eq.(3) has been completed.

\section{Ground state current in the presence of screw dislocation: analytical calculation}
In this section, we calculate the ground state current raised by the TCME in the case of screw dislocation,
by calculating directly the eigenstates of the Hamiltonian with the torsion.
This is an alternative approach for the derivation of the TCME, which does not rely on Eq. (3).
For this purpose, we, first, analyze
the spectrum of the Hamiltonian (Eq.(4) in the main text), 
\begin{align}
&H^{{\rm screw}}_{s,k_z}=\chi_{s} \left[  H^{\perp}_{k_z}  + m^{s}_{k_z} \sig^z\right], \nonumber \\
&H^{\perp}_{k_z}= \left(-\mathrm{i}\part_x- \frac{\Phi_{k_z} y}{2\pi \rho^2} \right)\sig^x + \left(-\mathrm{i}\part_y+ \frac{\Phi_{k_z}x}{2\pi \rho^2}\right)\sig^y.
\end{align}
where $m^{s}_{k_z}=k_z-\chi_{s}\lam$, $\Phi_{k_z}=k_zb_g$, $\rho=\sqrt{x^2+y^2}$, $\chi_{L(R)}=+1(-1)$, and $\sigma^i$ is the Pauli matrix.

For the calculation of the spectrum,
it is useful 
to clarify the symmetry of the eigenstates of $H^{\perp}_{k_z}$.
Suppose $\ket{\kappa}_{k_z}$ the eigenstate of $H^{\perp}_{k_z}$ with eigenvalue $\kappa$.
Since $\{H^{\perp}_{k_z},\sig^z\}=0$, where $\{U,V\}:=UV+VU$ is the anticommutator,the state $\sig^z\ket{\kappa}_{k_z}$ is also the eigenstate with eigenvalue $-\kappa$.
Therefore, we can choose the eigenfunctions to preserve the doublet structure, $\ket{-\kappa}_{k_z}=\sig^z\ket{\kappa}_{k_z}$,
for $\kappa\neq{0}$.
On the other hand, there is no double structure in the zero eigenstates.
Since the hermitian operator $\sig^z$ maps zero eigenstates of $H^{\perp}_{k_z}$ to zero eigenstates of $H^{\perp}_{k_z}$,
then we can choose the zero eigenstates also as eigenstates of $\sig^z$, denoted by $\ket{0_{i,\sig_i}}_{k_z}$ with $\sig^z\ket{0_{i,\sig_i}}_{k_z}=\sig_i\ket{0_{i,\sig_i}}_{k_z}$
and $\sig_i=\pm1$.
There is another symmetrical property between the eigenstates of $H^{\perp}_{k_z}$ with different $k_z$.
Since the transformation $k_z\to{-k_z}$ corresponds to the flip of the direction of the effective magnetic field,
$\Theta{H}^{\perp}_{k_z}\Theta^{-1}=H^{\perp}_{-k_z}$ holds,
where $\Theta=\mathrm{i}\sig^yK$ is the time-reversal operator for spin-$1/2$ fermions and $K$ is the complex conjugation operator \cite{Sakurai}.
Therefore we can impose $\ket{\kappa}_{-k_z}=\Theta\ket{\kappa}_{k_z}$
and $\ket{0_{i,-\sig_i}}_{-k_z}=\Theta\ket{0_{i,\sig_i}}_{k_z}$, because of $\{{\sig^z},\Theta\}=0$.

The eigenstates of $H^{{\rm screw}}_{s,k_z}$ can be constructed from $\ket{\kappa}_{k_z}$ and $\ket{0_{\sig_i}}_{k_z}$.
Indeed, $\ket{\psi^{L,\pm}_{k_z}(\kappa)}:=c^{L,\pm}_{k_z,1}(\kappa)\ket{\kappa}_{k_z}+c^{L,\pm}_{k_z,2}(\kappa)\ket{-\kappa}_{k_z}$, with $\kappa>0$,
and $\ket{0_{\sig_i}}_{k_z}$ are the full spectrum of $H^{{\rm screw}}_{L,k_z}$,
with eigenvalues 
$\pm \sqrt{\kappa^2 +(m^L_{k_z})^2}$ and $\sig_i m^L_{k_z}$, respectively.
Here the coefficients are given by $(c^{L,\pm}_{k_z,1}(\kappa),c^{L,\pm}_{k_z,2}(\kappa))=(4(\kappa^2+(m^L_{k_z})^2))^{-1/4}(\pm{\rm sgn}(m^L_{k_z})((\kappa^2+(m^L_{k_z})^2)^{1/2}\pm\kappa)^{1/2},((\kappa^2+(m^L_{k_z})^2)^{1/2}\mp\kappa)^{1/2}))$.
Moreover, $\ket{\psi^{R,\pm}_{k_z}(\kappa)}:=\Theta\ket{\psi^{L,\mp}_{-k_z}(\kappa)}$ and $\ket{0_{\sig_i}}_{k_z}$ are the eigenstates of $H^{{\rm screw}}_{R,k_z}$,
with eigenvalues 
$\pm \sqrt{\kappa^2 +(m^R_{k_z})^2}$ and $-\sig_i m^R_{k_z}$, respectively.

Now, we calculate the ground state current in the presence of the screw dislocation.
As yet, we have not distinguished discrete and continuum states.
From now on, we use $\kappa_i$ to express the discrete eigenvalues of $H^{\perp}_{k_z}$ and 
$(\kappa,l)$ to label the continuum states, where $\kappa$ is the continuum energy eigenvalue, and $l$ is a discrete quantum number, e.g., the angular momentum. 
The current operator is defined by $\part{H^{{\rm screw}}_{s,k_z}}/\part{k_z}=\chi_s\sig^z+\chi_s\{ -(b_gy/2\pi\rho^2)\sig^x+(b_gx/2\pi\rho^2)\sig^y \}$. At least up to the first order in $b_g$, the correction to the current operator due to the dislocation, i.e. the second and third term above, does not contribute to the expectation value 
because these terms are odd under the transformation
$x\to-x$ or $y\to-y$.
Therefore, the current is the sum of the expectation values of $\chi_s\sig^z$ 
for the occupied states which consist of discrete nonzero, discrete zero, and continuum states,
and then we obtain,
\begin{align}
J^z&= \sum_{s=L,R} \int_{|m^s_{k_z}|<\Lam} \frac{L_z dk_z}{2\pi} \left[ \sum_{\kappa_i > 0} \braket{\psi^{s,-}_{k_z}(\kappa_i)|\chi_s\sig^z|\psi^{s,-}_{k_z}(\kappa_i)} \right. \nonumber \\
&+\sum_{i: \sig_i \chi_s m^s_{k_z} <0}\braket{0_{i,\sig_i}|\chi_s\sig^z|0_{i,\sig_i}}_{k_z} \nonumber \\
&\left. + \int_0^{\infty} d\kappa \sum_l \braket{\psi^{s,-}_{k_z}(\kappa,l)|\chi_s\sig^z|\psi^{s,-}_{k_z}(\kappa,l)}\right] \nonumber \\
&=\int_{|m^L_{k_z}|<\Lam} \frac{L_z dk_z}{2\pi} \left[ \sum_{\kappa_i \neq 0} \braket{\kappa_i|\sig^z|\kappa_i}_{k_z} \right. \nonumber\\
&\left. + \sum_{i} \braket{0_{i,\sig_i}|\sig^z|0_{i,\sig_i}}_{k_z} +\int_{-\infty}^{\infty} d\kappa \sum_l \braket{\kappa,l | \sig^z|\kappa,l}_{k_z} \right], \label{current_screw_gs}
\end{align}
where $L_z$ is the size of the system.
Here we introduce the momentum cutoff scheme, $|m^s_{k_z}|<\Lam$,
i.e. the domain of the integration is 
the same as that used 
in the calculation of Eq.(4).
The first term in the square braket is equal to zero, since $\sig^z\ket{\kappa_i}_{k_z}=\ket{-\kappa_i}_{k_z}$ is orthogonal to $\ket{\kappa_i}_{k_z}$.
The second term is the index of the Dirac operator, $H^{\perp}_{k_z}$, which is an integer and the difference in the number of its normalizable zero modes with $\sig^3=+1$ and $\sig^3=-1$.
The index is given by $N_{k_z}:=-{\rm sgn}({\Phi_{k_z}})\lfloor{|\Phi_{k_z}|/2\pi}\rfloor$ \cite{Kiskis1977,Aharonov1979}.
The normalizable zero modes exhibit power-law decay for large distance from the dislocation; i.e. they behave like
$\ket{0_{i,-1}}\propto(0,\rho^{-\Phi_{k_z}/2\pi}(x-\mathrm{i}y)^{i-1})$ for $\Phi_{k_z}>0$, and 
$\ket{0_{i,+1}}\propto(\rho^{\Phi_{k_z}/2\pi}(x+\mathrm{i}y)^{i-1},0)$ for $\Phi_{k_z}<0$, where
$i=1,2,\cdots,|N_{k_z}|$ \cite{Kiskis1977,Aharonov1979}.
Now, we move on to the third term of Eq.(\ref{current_screw_gs}).
One may expect that it is equal to zero, since $\sig^z\ket{\kappa,l}=\ket{-\kappa,l}$, is orthogonal to $\ket{\kappa,l}$ for almost all values of $\kappa$.
However, the scattering states near $\kappa=0$ (their amplitudes $\propto \cos(\kappa\rho+\delta_l)/\sqrt{\rho}$ with $\delta_l$ the phase shift) cause 
a delta function peak of $\braket{\kappa,l | \sig^z|\kappa,l}_{k_z}$ at $\kappa=0$.
Indeed, from an explicit calculation \cite{Kiskis1977}, it has been shown that
$\sum_l \braket{\kappa,l | \sig^z|\kappa,l}_{k_z}=c_{k_Z}\delta(\kappa)$, with $c_{k_z}=\Phi_{k_z}/2\pi-N_{k_z}$,
and then the third term is equal to $c_{k_z}$.
Substituting them into Eq.(\ref{current_screw_gs}), we obtain
\begin{align}
J^z=\int_{-\Lam-\lam}^{\Lam-\lam} \frac{L_zdk_z}{2\pi}\frac{\Phi_{k_z}}{2\pi}=-\frac{L_zb_g\Lam\lam}{2\pi^2}, \label{current_ground_final}
\end{align}
which is coincident with the expression obtained directly from Eq.(3) by the following reason.
In the presence of the screw dislocation with the Burgers vector $-b_g\hat{z}$
the torsion is given by $T^z_z=T^z_{xy}=b_g\delta^{(2)}(x,y)$.
Therefore, the total current derived from Eq.(3) is $J^z=-L_zev_F(\lam^R_z - \lam^L_z)\Lam b_g/4\pi^2$.
In this section we have set $ \lam^L_z=-\lam^R_z =\lam$ and $e=v_F=1$,
and therefore we obtain $J^z=-L_z\lam \Lam b_g/2\pi^2$, which reproduces Eq.(\ref{current_ground_final}).

\section{Absence of total current and possibility of local current}
In this section, we show that, if the system is periodic in a certain direction, the {\it total} current is always zero,
even in the presence of a magnetic field or lattice stain and dislocations,
while the {\it local} current is not.
The argument is the extension of that presented in Ref. \cite{Vazifeh2013}.
We start with the general Hamiltonian of electrons in solids (set $e=1$ in this section):
\begin{align}
H=\int d^3{\bm r} \ \frac{1}{2m} \left( -\mathrm{i} \nab_i - A_i({\bm \rho},z) \right)^2 + V({\bm \rho} ,z),
\end{align}
where $V$ is the potential term, in which the effect of the dislocation is included.
Here $i=x,y,z$, ${\bm r}=({\bm \rho},z)$, and ${\bm \rho}=(x,y)$, and
we impose the periodicity in the $z$-direction:
\begin{align}
A_i({\bm \rho}, z) = A_i({\bm \rho}, z+a), V({\bm \rho} ,z) = V({\bm \rho} ,z+a).
\end{align}
Suppose $\psi$ is one eigenfunction of the Hamiltonian and define the Bloch wave function
$\psi_{n,k_z}({\bm \rho}, z) = e^{\mathrm{i} k_z z}u_{n,k_z}({\bm \rho}, z)$, whose energy is $\vareps_{n,k_z}$.
The total current along the $z$-direction is given by
\begin{align}
J_z
 =   \sum_{n} \int_{BZ} \frac{d k_z}{2 \pi} \int d^3 {\bm r}  \ \psi^*_{n,k_z}({\bm \rho},z)\frac{\delta H}{\delta A_z} \psi_{n,k_z}({\bm \rho},z) n_F(\vareps_{n.k_z})\nonumber \\
=-  \sum_{n} \int_{BZ} \frac{d k_z}{2 \pi} \int d^3 {\bm r}  \ u^*_{n,k_z}({\bm \rho},z)\frac{\part H_{k_z}}{\part k_z} u_{n,k_z}({\bm \rho},z) n_F(\vareps_{n.k_z}), \nonumber \\
\label{jay-z_total}
\end{align}
where $H_{k_z} = e^{-i k_z z} H e^{i k_z z} $ and $n_F$ is the Fermi distribution function.
Here we use the identity :
\begin{align}
&\int d^3 {\bm r}  \ u^*_{n,k_z}({\bm \rho},z)\frac{\part H_{k_z}}{\part k_z} u_{n,k_z}({\bm \rho},z) \nonumber \\
& = \frac{\part}{\part k_z} \int d^3 {\bm r}  \ u^*_{n,k_z}({\bm \rho},z)   H_{k_z} u_{n,k_z}({\bm \rho},z) \nonumber \\
& = \frac{\part \vareps_{n,k_z}}{\part k_z},
\end{align} 
which follows from
\begin{align}
\frac{\part}{\part k_z}  \left[ \int d^3 r  \ u^*_{n,k_z}({\bm \rho},z) u_{n,k_z}({\bm \rho},z) \right] = \frac{\part}{\part k_z} 1 =0,  \label{der_zero}
\end{align}
and then we can rewrite Eq. (\ref{jay-z_total}) into
\begin{align}
J_z&= - \sum_{n} \int_{BZ} \frac{d k_z}{2 \pi}  \frac{\part \vareps_{n,k_z}}{\part k_z} n_F(\vareps_{n.k_z}) \nonumber \\
&=  - \frac{1}{2 \pi}  \sum_{n}  \sum_{i = 1,\dots ,i^{(n)} } \int_{\vareps_{n, k^{(n)}_{i-1}}}^{ \vareps_{n, k^{(n)}_{i}}} d \vareps \  n_F( \vareps)  \label{jay-z-eps}
\end{align}
Here for each region $k \in (k^{(n)}_{i-1}, k^{(n)}_{i})$, $\vareps_{n,k}$ monotonically increases or decreases, 
and $k^{(n)}_0=0$ and $k_{i^{(n)}}^{(n)} = 2\pi/a$.
We find that Eq.(\ref{jay-z-eps}) is always equal to zero owing to the periodicity of the dispersion in the wave number space, $\vareps_{n,k=0}=\vareps_{n, k=2\pi/a}$.
Therefore, we found that the {\it total} current along the $z$-direction is zero.
In the above derivation, 
it is essential that the integrand with respect to $k_z$ can be rewritten into the total derivative with respect to $k_z$,
and this key factor follows from the fact that the integral over the real space of $|u_{n,k_z}({\bm \rho},z)|^2$ is equal to 1, which results 
in Eq. (\ref{der_zero}).
Instead, without the integration over the real space,
\begin{align}
 \frac{\part }{\part k_z} \left[  u^*_{n,k_z}({\bm \rho},z) u_{n,k_z}({\bm \rho},z) \right]  \neq 0.
\end{align}
Then, in the case of {\it local} current, the above argument 
in the case of the {\it total} current does not hold.
Hence, the local current is not always zero, unlike the total current.

\bibliography{WSM}
\bibliographystyle{apsrev}

\end{document}